\documentclass[prl,aps,twocolumn,longbibliography]{revtex4-1}
\usepackage{color}
\usepackage{graphicx}
\usepackage{bm}
\usepackage{amsmath}
\usepackage{amssymb}
\usepackage{amsthm}
\usepackage{amsfonts}
\usepackage{enumerate}
\usepackage{latexsym}
\usepackage{hyperref}

\begin{document}

\title{Curved spacetime theory of inhomogeneous Weyl materials}
\author{Long Liang$^{1,2}$ and  Teemu Ojanen$^{2}$}
\affiliation{1, Department of Applied Physics, Aalto University School of Science, FI-00076 Aalto, Finland \\2, Computational Physics Laboratory, Physics Unit, Faculty of Engineering and
Natural Sciences, Tampere University, P.O. Box 692, FI-33014 Tampere, Finland}

\begin{abstract}
We show how the universal low-energy properties of Weyl semimetals with spatially varying time-reversal (TR) or inversion (I) symmetry breaking are described in terms of chiral fermions experiencing curved-\emph{spacetime} geometry and synthetic gauge fields. By employing Clifford representations and Schrieffer-Wolff transformations, we present a systematic derivation of an effective curved-space Weyl theory with rich geometric and gauge structure. To illustrate the utility of the formalism, we give a concrete prescription of how to fabricate nontrivial curved spacetimes and event horizons in topological insulators with magnetic textures. Our theory can also account for strain-induced effects, providing a powerful unified framework for studying and designing inhomogeneous Weyl materials. 

\end{abstract}
\maketitle

\emph{Introduction-- }Semimetals and quantum liquids with linear dispersion near degeneracy points exhibit emergent relativistic physics at low energies. Topological Dirac and Weyl semimetals~\cite{Volovik:2003,Shuichi_Murakami_2007,PhysRevLett.108.140405,PhysRevB.85.195320,PhysRevB.88.125427,2014Sci...343..864L,2014NatCo...5E3786N,Liu2014,PhysRevB.83.205101,PhysRevLett.113.027603,PhysRevLett.107.186806,PhysRevLett.107.127205,PhysRevB.84.235126,PhysRevB.85.035103,PhysRevX.5.031013,PhysRevX.5.031023,2015Sci...349..613X,PhysRevLett.115.265304,Soluyanov2015} have proven particularly fertile condensed-matter playground to study the interaction of chiral fermions with gauge fields. These systems display the rich physics of quantum anomalies discovered originally in the relativistic setting~\cite{PhysRev.177.2426,Bell1969,1983PhLB..130..389N}. In translationally invariant systems the two-fold band-touching in Weyl semimetals can be associated with a conserved Berry charge which is topologically protected. Moreover, even in the presence of spatially varying perturbations the low-energy properties can be understood in terms of Weyl particles experiencing artificial gravity and gauge fields~\cite{WeylMetamaterials_PRX2017}. The condensed-matter setting allows for remarkable opportunities in engineering synthetic gauge fields and geometries that mimic and generalize the phenomenology of high-energy physics~\cite{PhysRevB.87.235306,PhysRevB.88.104412,PhysRevD.88.025040,PhysRevX.5.031023,2016PNAS..113.9463L,2016NatCo...710735Z,PhysRevLett.116.166601,PhysRevB.96.161115,2017Natur.547..324G,Khaidukov2018,PhysRevB.99.140201,PhysRevLett.122.056601,2017Natur.547..324G}. 

A popular starting point for geometry and gauge-field engineering in semimetals is a strain-distorted tight-binding model \cite{PhysRevLett.108.227205,PhysRevB.87.165131,PhysRevLett.115.177202,PhysRevB.92.245110,PhysRevX.6.041046,PhysRevX.6.041021}. This description, case-specific to a particular lattice and orbital structure, can often be regarded as a formal device to obtain a long-wavelength theory. While being a powerful method for fabricating synthetic gauge fields, strain engineering has the limitation of producing effective geometries that are small perturbations from flat space. To obtain more general 3d geometries, Ref.~\onlinecite{WeylMetamaterials_PRX2017} proposed a new method of fabricating spatially varying TR and I-breaking textures. Semiclassical dynamics of carriers then reflect the interplay of effective curved geometry and Berry curvature effects. The purpose of the present work is to establish a general and fully quantum-mechanical description of inhomogeneous Weyl semimetals. By employing Clifford representations and Schrieffer-Wolff (SW) transformations~\cite{Winkler}, we systematically consider generic TR and I breaking patterns and provide a controlled derivation of the low-energy Weyl Hamiltonian 
$H_W=V\sigma^0+e^{i}_{~a}(k_i-K_{W,i}-\frac{i}{2}e^{~b}_{i}\partial_{j}e^{j}_{~b})\sigma^a.$ Here $\sigma^a$ denotes the set of Pauli matrices supplemented by a unit matrix. The effective geometry is encoded in the frame fields  $e^{i}_{~a}$ while the effective gauge field receives contributions from the spatial variation of the Weyl point $K_{W,i}$ and the frame fields. Our theory can also account for strain-induced effects and provides a unified low-energy description of inhomogeneous Weyl semimetals.

The obtained low-energy theory has a number of remarkable consequences. In general, spatially varying TR and I-breaking textures give rise to frame fields and metric tensors that mix time and space components. In contrast to mere curved \emph{space} geometries realized by strain engineering, we obtain nontrivial \emph{spacetime} geometries. Also, the spectral tilt of the Weyl dispersion can be tuned by TR and I breaking textures. To demonstrate this effect in detail, we consider 3d topological insulators with magnetic textures. Strikingly, various magnetic textures give rise to Weyl semimetal phase with a spatial interface between type I and type II regions. We show that the effective geometry near the interface emulates the Schwarzschild metric at a black hole event horizon. Our work provides powerful tools to analyze and design the properties of inhomogeneous Weyl materials.

\emph{Inhomogeneous Weyl systems-- } 
The starting point of our theory of inhomogeneous Weyl systems is a generic four-band parent state with both time-reversal and inversion symmetry intact. By introducing spatially varying TR or I breaking fields the parent state is driven to an inhomogeneous Weyl semimetal phase. The minimal model for the parent states 
is  characterized by the Hamiltonian~\cite{WeylMetamaterials_PRX2017}
\begin{eqnarray}
H_0=n(\mathbf{k})\mathbb{I} +\kappa_i(\mathbf{k}) \gamma_i + m(\mathbf{k}) \gamma_4,
\end{eqnarray}
where the repeated indices are implicitly summed over, 
$\mathbb{I}$ is the identity matrix, and $\gamma_\mu$ with $\mu=1,2,3,4$ denotes the four gamma matrices satisfying anticommutation relations $\{\gamma_{\mu},\gamma_{\nu}\}=2\delta_{\mu\nu}$. The parameters
$n(\mathbf{k})$ and $m(\mathbf{k})$ are even functions of the momentum while $\kappa_i(\mathbf{k})$ is odd. Therefore $\gamma_{1,2,3}$ are odd under both TR and I, while $\gamma_4$ is even. 
The fifth gamma matrix is defined as $\gamma_5=\gamma_1\gamma_2\gamma_3\gamma_4$, which is odd under both TR and I.

To write down the most general $4\times 4$ Hamiltonian we introduce ten additional matrices, $\gamma_{ij}=-i[\gamma_i,\gamma_j]/2$.  It is convenient \cite{PhysRevB.84.235126,WeylMetamaterials_PRX2017} to separate the ten matrices into three vectors  $\mathbf{b}=(\gamma_{23},\gamma_{31}, \gamma_{12})$, $\mathbf{b}'=(\gamma_{15},\gamma_{25},\gamma_{35})$, $\mathbf{p}=(\gamma_{14},\gamma_{24},\gamma_{34})$, and one scalar $\varepsilon=\gamma_{45}$. The transformation properties of the four groups can be easily deduced from their constituent gamma matrices, and it turns out that $\mathbf{b}$ and $\mathbf{b}'$ break TR symmetry, while $\mathbf{p}$ and $\varepsilon$ break I symmetry.
The general $4\times 4$ Hamiltonian with TR and I breaking terms can thus be written as
\begin{eqnarray}\label{Eq:TRIB_Dirac}
H=H_0+\mathbf{u}\cdot\mathbf{b}+\mathbf{w}\cdot\mathbf{p}
+\mathbf{u}'\cdot\mathbf{b}'+f\varepsilon,
\end{eqnarray}
where the functions $\mathbf{u}$, $\mathbf{w}$, $\mathbf{u}'$, and $f$ characterize the symmetry breaking fields which are position dependent. Because $\mathbf{r}$ is even under TR but odd under I, $\mathbf{w}$ and $f$ should be even functions of $\mathbf{r}$ to break the inversion symmetry. If $\mathbf{u}$ and $\mathbf{u}'$ are not even functions of $\mathbf{r}$, the inversion symmetry is also broken. But since $\mathbf{u}$ and $\mathbf{u}'$  fields always break time-reversal symmetry irrespective of their $\mathbf{r}$ dependence, we label them as time reversal symmetry breaking terms. In general, we regard TR and I-breaking terms as smooth functions position $\mathbf{r}$ and assume that $\mathbf{r}$ and $\mathbf{k}$ are conjugate variables  $[r_i,k_j]=i\delta_{ij}$. Elastic deformations and strain typically induce spatial dependence in $H_0$~\cite{PhysRevLett.108.227205,PhysRevB.87.165131,PhysRevLett.115.177202,PhysRevB.92.245110,PhysRevX.6.041046,PhysRevX.6.041021}. While we are mainly considering inhomogeneous TR and I breaking, we will discuss below how strain is included in our formalism in the continuum limit.

\emph{TR-breaking case-- } Here we derive an effective curved-space Weyl equation for inhomogeneous TR breaking systems in two opposite limits, vanishing mass ($m=0$) and large mass. Physically these correspond to metallic and insulating parent states. We consider $\mathbf{u}(\mathbf{r})\cdot\mathbf{b}$ term which corresponds to 3d magnetization or any field which transforms as magnetization under TR and spatial rotations. The unit matrix term $n(\mathbf{k})$ can be set zero since it only shifts the energy. For the $m(\mathbf{k})=0$ case the Hamiltonian can be readily block diagonalized. Defining two sets of Pauli matrices $\sigma_i$ and $\tau_i$ ($\sigma_0$ and $\tau_0$ are the $2\times 2$ unit matrix) and working in the chiral representation $\gamma_i=\tau_3\otimes \sigma_i$, $\gamma_{4}=\tau_1\otimes \sigma_0$, the Hamiltonian splits to two blocks
\begin{eqnarray}
H^{\pm}_W=[u_i(\mathbf{r})\pm \kappa_i(\mathbf{k}) ] \sigma_i=d_i^{\pm}(\mathbf{k},\mathbf{r})\sigma_i.
\end{eqnarray}
The local Weyl point are determined by $d_i^{\pm}=u_i(\mathbf{r})\pm \kappa_i(\mathbf{K}_{W})=0$. Thus $\mathbf{u}(\mathbf{r})$ give rise to an axial gauge field while the frame fields can be straightforwardly obtained as indicated [by Eq.~\eqref{frame}] below.  

The complementary regime of nonzero mass gives rise to  rich geometric and gauge structure.  It is convenient to parametrize the symmetry breaking fields as $\mathbf{u}(\mathbf{r})=u(\mathbf{r})(\sin{\theta(\mathbf{r})}\cos{\phi(\mathbf{r})}, \sin{\theta(\mathbf{r})}\sin{\phi(\mathbf{r})}, \cos{\theta(\mathbf{r})})$.
To derive an effective Weyl Hamiltonian, we first rotate the fields along the $z$ direction by applying a unitary transformation, $W^\dag(\mathbf{r}) \mathbf{u}\cdot\mathbf{b} W(\mathbf{r})=u(\mathbf{r}) b_3$. This is achieved by choosing $W=\exp{(-i\phi b_3/2 )}\exp{(-i\theta b_2/2)}$.
%To derive an effective Weyl Hamiltonian in this case, we first rotate the fields along the $z$ direction by applying a unitary transformation, $W^\dag(\mathbf{r}) \mathbf{u}\cdot\mathbf{b} W(\mathbf{r})=u(\mathbf{r}) b_3$ where $u(\mathbf{r})=|\mathbf{u}|$. 
The Hamiltonian after the transformation becomes
\begin{eqnarray}\label{Eq:TRB_Dirac}
 H' %&=&W^\dag H W %=W^\dag H_0 W+h(\mathbf{r})\gamma_{12},\\
&=&\frac{1}{2}\{ E^i_{~a}\gamma^a, \kappa_i(\mathbf{k} -\bm{\omega}) \} +m(\mathbf{k}-\bm{\omega})\gamma_4+u\gamma_{12}.~~
\end{eqnarray}
The anticommutator structure results from the noncommutativity of position and momentum. Here $E^i_{~a}$ with $i$ and $a$ running from 1 to 3 are frame fields defined through $W^\dag \gamma_i W=E^i_{~a}\gamma^a$. Using the frame fields we can define a metric  $g^{ij}=E^i_{~a}E^j_{~b}\eta^{ab}$ with $\eta$ being the Minkowski metric, $\eta=\mathrm{diag}(-1,1,1,1)$. The indices $i,j,k$ can be raised or lowered by $g^{ij}$ or its inverse $g_{ij}$, and $a,b,c$ are raised or lowered by  $\eta^{ab}$ or $\eta_{ab}$. It is easy to verify that $E^i_{~a} E_i^{~b}=\delta^b_a$ and $E^i_{~a} E_j^{~a}=\delta^i_j$.
 The spin connection
$\bm{\omega}=i W^\dag \partial_{\mathbf{r}} W$  can be written in terms of the frame fields as $\omega_i = \omega^{~ab}_i\gamma_{ab} =- E^{~a}_j \nabla_i E^{j b} \gamma_{ab}/4 $, with $\bm{\nabla}$ being the covariant derivative on the manifold.
Alternatively, $\bm{\omega}$ can also be viewed as a $SU(2)$  gauge connection $\omega_i=\omega_{i}^{~j}b_j$. 
Eq.~\eqref{Eq:TRB_Dirac} describes a Dirac electron moving in a curved space~\cite{2011arXiv1106.2037Y} in the presence of time reversal symmetry breaking field $u$. The elegant gauge structure of Eq.~\eqref{Eq:TRIB_Dirac} is discussed in detail in the Supplemental Information (SI). While the Dirac equation still describes flat space $g^{ij}=\delta^{ij}$, it contains redundant high-energy bands. The curved space geometry and gauge fields emerge as we project the four band Dirac Hamiltonian, Eq.~\eqref{Eq:TRB_Dirac}, to the two band low-energy Weyl Hamiltonian. If the symmetry breaking field $\mathbf{u}$ is constant, this can be done exactly by a momentum dependent unitary transformation. However, for position dependent fields we are interested in, this method is no longer exact since the momentum and position do not commute with each other. There will be additional terms related to the derivatives of the fields that mix the high and low energy degrees of freedom, and in general it is impossible to perform the block diagonalization. However, in the large mass limit and slowly varying fields we can derive an effective low energy theory in a controlled way by employing the SW transformation. 

First we expand Eq.~\eqref{Eq:TRB_Dirac} to the first order in derivatives of the symmetry breaking field. This yields 
\begin{eqnarray}\label{intermediate}
H'\approx \tilde{\kappa}_i(\mathbf{k},\mathbf{r})\gamma_i+m(\mathbf{k})\gamma_4+u\gamma_{12}+u'_{i} b'_i+f\gamma_{45},
\end{eqnarray}
where $\tilde{\kappa}_i=\{\kappa_i(\mathbf{k}), E^i_{~a}(\mathbf{r})\}/2$, $u'_i=\partial_{k_j} m\omega_j^{~i}$,  and $f=\partial_{k_j}\kappa_i\omega_j^{~a}E^i_{~a}$.
Choosing a particular representation $\gamma_1=\tau_0\otimes\sigma_x$, $\gamma_2=\tau_0\otimes\sigma_y$, $\gamma_3=\tau_x\otimes\sigma_z$, $\gamma_4=\tau_z\otimes\sigma_z$, the terms proportional to $\gamma_{1,2,4}$, $b_3$ and $b_3'$ are block diagonal. Employing the SW transformation, we seek matrix $S$ such that the unitary equivalent Hamiltonian $e^SH'e^{-S}$ is block diagonal. In the SI we explicitly write $S$ in the lowest order in the large mass limit $u'_i/m,\kappa_3'/m\ll 1$.  The transformed Hamiltonian is block-diagonal, yielding an effective Weyl Hamiltonian
\begin{eqnarray} \label{Eq:twoband}
H_W=d_a(\mathbf{k},\mathbf{r})\sigma^a,
\end{eqnarray}
with $d_{i=1,2}=\tilde{\kappa}_i-f u'_{i}/(2m)-f u'_{i}/(2u)$, $d_3=-m+u-(\tilde{\kappa}^{2}_3+f^{2})/(2m)+(u^{'2}_{1}+u^{'2}_{2})/(2u)$, and $d_0=u'_{3}+f \tilde{\kappa}_3/m$.
The Weyl points are determined by $d_{i=1,2,3}=0$, and due to the inversion symmetry, if $\mathbf{K}_W$ is a Weyl point, $-\mathbf{K}_W$ will also be a Weyl point. The higher-order corrections to Eq.~\eqref{Eq:twoband} are proportional to $\partial_{k}^2\kappa_i(\partial_{r}\mathbf{u})^2$, $\partial_{k}^2 m(\partial_{r}\mathbf{u})^2$, which are always small for smooth $\mathbf{u}$ and vanish completely for $H_0$ with a linear dispersion. Expanding the Hamiltonian around $\mathbf{K}_W$, we arrive at our main result
\begin{eqnarray}\label{Eq:Weyl}
H_W\approx V\sigma^0+ e^{i}_{~a}(k_i-K_{W,i}-\frac{i}{2}e^{~b}_{i}\partial_{j}e^{j}_{~b})\sigma^a,
\end{eqnarray}
where $i=1,2,3$ and $a=0,1,2,3$.
The Weyl point depends on the position and can be interpreted as $U(1)$ vector potential, and
$V=d_0(\mathbf{K}_W)-\frac{i}{2}e^{~b}_0\partial_j e^j_{~b}$ acts as an effective scalar potential.  The potentials acquire corrections from the frame fields, ensuring the Hamiltonian Hermitian. The frame fields  are defined through
\begin{eqnarray} \label{frame}
 e^i_{~a}(\mathbf{r})=\frac{\partial d_a}{\partial k_i}\bigg|_{\mathbf{K}_W},
\end{eqnarray}
with additions $e^{0}_{~0}=-1$  and $e^{0}_{~1,2,3}=0$.
We obtain from the frame fields the emergent metric
$g^{\mu\nu}=e^{\mu}_{~a}e^{\nu}_{~b}\eta^{ab}$. Explicitly,
\begin{eqnarray}
g^{00}=-1,~g^{0i}=e^{i}_{~0},~g^{ij}=-e^i_{~0}e^j_{~0}+e^i_{~a}e^j_{~a}.
\end{eqnarray}
It should be stressed that, as indicated by the effective metric tensor with space and time mixing terms $g^{0i}$, the emergent geometry of Eq.~\eqref{Eq:Weyl} is fundamentally different from the inhomogeneous strain induced metrics that do not contain the mixing terms. The dispersion relation of the Weyl fermion can be determined by $g^{\mu\nu}p_\mu p_\nu=0$  with $p=(\omega,\mathbf{k})$, which gives 
\begin{eqnarray}
\omega=e^i_{~0}k_i\pm\sqrt{e^i_{~l}e^j_{~l}k_i k_j}.
\end{eqnarray}
%$\omega=[-g^{0i}k_i\pm\sqrt{(g^{0i}k_i)^2-g^{00}g^{ij}k_i k_j }]/g^{00}=e^i_{~0}k_i\pm\sqrt{e^i_{~l}e^j_{~l}k_i k_j}$.
 If $|e^{i}_{~0}e^{~l}_i|<1$, it is a type I Weyl semimetal, and $|e^{i}_{~0}e^{~l}_i|>1$ we obtain a type II Weyl semimetal with over-tilted Weyl cone. Remarkably, as shown below, this fact can be employed in engineering a spatial interface between type I and type II Weyl semimetals. The interface between type I and type II Weyl semimetals, which could simulate properties of the black hole horizon, may be designed experimentally by controlling the magnetic texture. This could open a method to study  Hawking radiation~\cite{2016JETPL.104..645V,Volovik2017} and quantum chaos~\cite{Shenker2014,Kitaevtalks,Maldacena2016,2019arXiv190310886C} in Weyl semimetals.

To conclude this section, we outline how things would have changed had we considered the other TR-breaking triplet $\mathbf{u}'\cdot \mathbf{b}'$ instead of $\mathbf{u}\cdot \mathbf{b}$. In this case, we can use essentially the same method to derive a two-band model, but in general this leads to a nodal line semimetal~\cite{PhysRevB.84.235126} instead of a Weyl semimetal. While interesting, this case is not relevant for Weyl-type behaviour and postponed to the SI.

\emph{I-breaking case-- } The derivation of the curved-space Weyl Hamiltonian can be extended to  I-breaking systems. Several Weyl materials that break inversion (but preserve TR) symmetry have be observed~\cite{PhysRevX.5.031013,PhysRevX.5.031023,2015Sci...349..613X}. In general, a spatially varying I-breaking term $\mathbf{w}(\mathbf{r})\cdot\mathbf{p}$ can be treated similarly as the TR breaking case. This has been carried out in the SI where we obtain an I-breaking variant of Eq.~\eqref{Eq:twoband}. 
The essential difference from the TR-breaking case is that the role of the mass term $m$ is played by $\tilde{\kappa}_1$ or $\tilde{\kappa}_2$. Thus, the controlling parameter in this situation is $\tilde{\kappa}_1$ (or $\tilde{\kappa}_2$) instead of $m$. Assuming $\tilde{\kappa}_1$ sufficiently large, which is also a required to obtain Weyl points in this case, we can find a SW transformation to block-diagonalize $H'$ and obtain a two-band Hamiltonian.
Finally, analogously to the $\mathbf{u}'\cdot\mathbf{b}'$ term, the $f\gamma_{45}$ term as a sole I-breaking term leads to a nodal line semimetal as shown in the SI.

\emph{Application I: engineering tilts and horizons--}  Here we illustrate the power and utility of the developed formalism by proposing concrete systems with spatial interface between type I and type II Weyl fermions. We consider a simple parent model with $\kappa_i=k_i$ and $m(\mathbf{k})=m$ which could be realized in 3d TR-invariant insulators with magnetization texture ~\cite{WeylMetamaterials_PRX2017} or in the topological insulator\textendash magnet heterostructures~\cite{PhysRevLett.107.127205}. The TR breaking field, representing for example 3d magnetization, is parametrized as $\mathbf{u} 
=u(\mathbf{r}) (\sin{\theta}\cos{\phi},\sin{\theta}\sin{\phi},\cos{\theta})$ with spatially varying angles. 

\begin{figure}
	\includegraphics[width=0.45\textwidth]{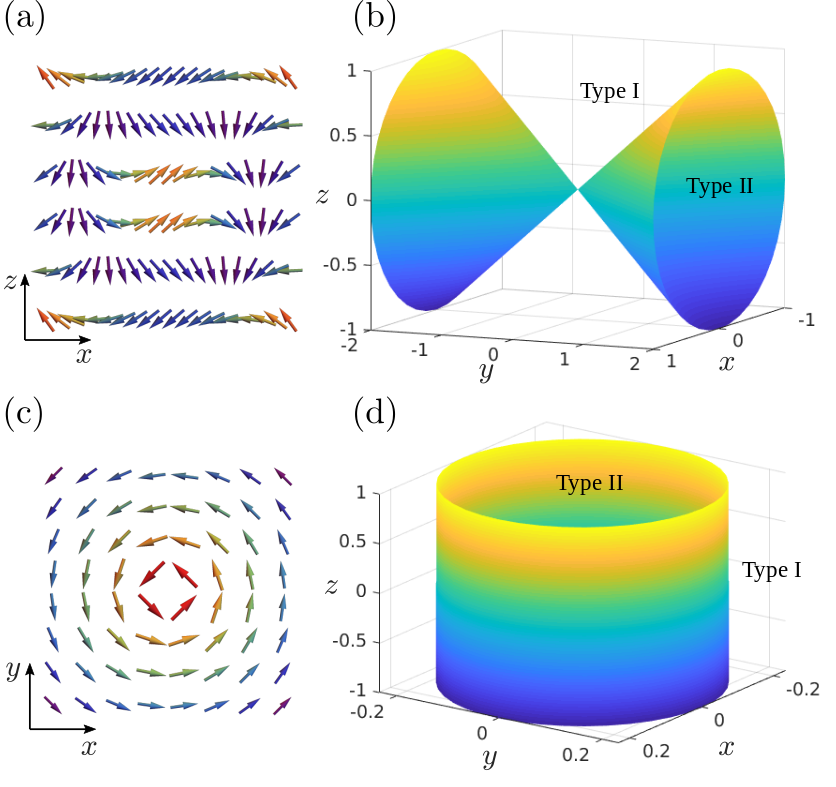}
	\caption{Magnetic textures and the corresponding event horizons.
	(a)-(b), $\mathbf{u}=u (\sin{r/\xi},0,\cos{r/\xi})$ with  $u/m=11/10$ and $m\xi=\sqrt{2}$. The horizon is a conical surface defined by $x^2+z^2=y^2/4$. 
 	(c)-(d), $\mathbf{u}=(-u_{||}y/\sqrt{x^2+y^2},u_{||}x/\sqrt{x^2+y^2} ,u_3)$, where $u_{||}=u_0 \exp{(-\sqrt{x^2+y^2}/\xi)}$, $u_3/m=10/9$, $u_0/u_3=1/9$, and $m\xi=9/70$. The horizon is a cylinder surface with the radius about $0.232\xi$.
 	The values of the coordinates in both plots are given in the units of $\xi$.}\label{Fig:interface}
\end{figure}

 Following the procedure discussed above, we obtain an effective $2\times2$ Hamiltonian with $d_1=k_x \cos{\theta}\cos{\phi} + k_y \cos{\theta}\sin{\phi} -k_z \sin{\theta}$,
$d_2=-k_x\sin{\phi}+k_y \cos{\phi}$, 
$d_3=u-m-(\tilde{\kappa}^{2}_3+f^2)/(2m)$, and $d_0=-\tilde{\kappa}_3 f/m$, where  $\tilde{\kappa}_3=k_x \cos{\phi}\sin{\theta}+k_y \sin{\phi}\sin{\theta}+k_z \cos{\theta}$ and $f=(\partial_{r_z}\phi + \cos{\phi}\partial_{r_y}\theta-\sin{\phi}\partial_{r_x}\theta)/2$. The products of $\mathbf{k}$ and $\mathbf{r}$ dependent terms should be understood as symmetrized. For the lowest-order of the SW transformation to be a good approximation around the Weyl point, the condition $u-m\ll m$ should be satisfied.  The Weyl points are determined by $\pm \mathbf{K}_W=\pm K_W(\sin{\theta}\cos{\phi}, \sin{\theta}\sin{\phi},\cos{\theta})$ with $K_W=\sqrt{2m(u-m)-f^2}$. For simplicity we assume that $2m(u-m)-f^2>0$ such that there always exist two well-separated Weyl points.
Expanding around $\mathbf{K}_W$, we obtain the linearized Weyl Hamiltonian with the frame fields being $e^0_{~0}=-1$, $e^i_{~0}=-f/m (\sin{\theta}\cos{\phi},\sin{\theta}\sin{\phi},\cos{\theta})$,
$e^i_{~1}=(\cos{\theta}\cos{\phi},\cos{\theta}\sin{\phi},-\sin{\theta})$, $e^i_{~2}= (-\sin{\phi},\cos{\phi},0)$, and $e^i_{~3}=-K_W/f e^i_{~0}$.  After straightforward calculations, we find $e^{i}_{~0}e^{~x}_i=e^{i}_{~0}e^{~y}_i=0$, and $e^{i}_{~0}e^{~z}_i=f/K_W$. Therefore 
if $|f/K_W|<1$, the node is of type I, if $|f/K_W|>1$, it is of type II, and the interface between the type I and type II regions, i.e., the event horizon, is determined by $|f/K_W|=1$. In the SI it is shown that in a suitable local basis the metric is analogous to the  Schwarzschild metric near the horizon in the Gullstrand-Painlev\'{e} coordinates~\cite{2016JETPL.104..645V,PhysRevLett.46.1351,PhysRevD.51.2827}.

Having worked out the general case, we now study a special case with $u$ as a constant, $\phi=0$, and $\theta=r/\xi$ with $r=\sqrt{x^2+y^2+z^2}$. Correspondingly, we find $f=y/(2r \xi)$. This magnetic texture is slowly varying in the length scale much smaller than $\xi$.   
Since $0\le |f|\le 1/(2\xi)$, there always exists a type I region, and to have a type II region, the condition $4m(u-m)\xi^2 < 1$ should be satisfied. The the event horizon is determined by $y^2=4\xi^2m(u-m)r^2$, which defines a conical surface, see Fig.~\ref{Fig:interface}(a)-(b). 
Clearly, the interface can be tuned when it is possible to manipulate $\xi$. Thus, the horizon may be tuned experimentally, providing a way to simulate the Hawking radiation using Weyl semimetals~\cite{2016JETPL.104..645V,Volovik2017}. As depicted by  Figs.~\ref{Fig:interface}(c)-(d),
 different shapes of the event horizon can also be realized, see SI for more details.
%In the SI we provide another texture that will give rise to type I-type II interface in magnetic topological insulators.
%\begin{widetext}

\emph{Application II: inclusion of strain effects-- }As noted above, there exists a significant body of literature on strain-induced artificial gauge fields and synthetic geometry \cite{PhysRevLett.108.227205,PhysRevB.87.165131,PhysRevLett.115.177202,PhysRevB.92.245110,PhysRevX.6.041046,PhysRevX.6.041021}. In these treatments the strain effects manifest at the level of the TR and I preserving parent state Hamiltonian $H_0$ which becomes position dependent. In this respect, the strain engineering can be viewed as complementary to the studied case with spatially dependent TR and I-breaking fields. However, here we show how to incorporate the strain effects to a unified theory of inhomogeneous Weyl systems in the continuum limit. In the presence of small strain, the spatial metric is related to the strain tensor $u_{ij}$ as $g_{ij}=\delta_{ij}+2u_{ij}$, and the frame fields can be chosen as  $\bar{E}^{i}_{~a}=\delta^{i}_a-\delta_{aj}u^{ij}$~\cite{PhysRevLett.108.227205}. Employing these frame fields $\bar{E}^{i}_{~a}$ and the corresponding spin connection $\bar{\bm{\omega}}$, the four band Dirac Hamiltonian in the presence of strain and TR breaking field $\mathbf{u}(\mathbf{r})\cdot\mathbf{b}$ can be written in a Hermitian form as (we assume $\kappa_i=k_i$ and $m(\mathbf{k})=m$)
\begin{eqnarray}
H_0=\frac{1}{2}\left\{\bar{E}^i_{~a}\gamma^a, k_i -\bar{\omega}_i \right\}+m\gamma_4+\mathbf{u}(\mathbf{r})\cdot\mathbf{b}.
\end{eqnarray}
This prescription can be understood as a momentum-dependent minimal substitution~\cite{PhysRevD.88.025040} accounting for strain. The strain-induced gauge fields~\cite{PhysRevLett.115.177202,PhysRevB.92.165131} are encoded in the frame fields and the spin connection~\cite{PhysRevLett.108.227205,2010PhR...496..109V}.
Using the same unitary transformation $W$ as previously, we can rotate $\mathbf{u}\cdot\mathbf{b}$ along the $z$ direction, and the Hamiltonian after rotation becomes
\begin{eqnarray}
H'&=&\frac{1}{2}\left\{\tilde{E}^i_{~a} \gamma^a, k_i -\tilde{\omega}_i \right\}+m\gamma_4 + u(\mathbf{r})b_3,\label{Eq:Strain_Dirac}
\end{eqnarray}
where $\tilde{E}^i_{~a}=\bar{E}^i_{~b} E^b_{~a}$
and  $\tilde{\bm{\omega}}=W^\dag \bar{\bm{\omega}} W+\bm{\omega}$ are the new frame fields and the modified spin connection containing the combined effects of strain and inhomogeneous TR breaking. In the SI we show that the spin connection transforms exactly in agreement with the modification of the frame fields. Thus, after inclusion of strain effects, Eq.~\eqref{Eq:Strain_Dirac} takes mathematically the same form as
Eq.~\eqref{Eq:TRB_Dirac} without strain. We can proceed with the projection to the low-energy space precisely as before to obtain an effective Weyl Hamiltonian of form Eq.~\eqref{Eq:Weyl}, now accounting for the strain and inhomogeneous TR-breaking texture.

%\begin{itemize}
%    \item {We should demonstrate the utility of the formalism by some examples. Simpler the better}
%     \item {For example, transition from type I to type II due to magnetic texture?}
%     \item {Explicit solution for the gauge fields in some simple rotating magnetic texture?}
%     \item {Discussion of semiclassical dynamics? This could be very illuminating for the effective geometry}
%\end{itemize}

\emph{Summary and outlook-- }
By employing Clifford representations and Schrieffer-Wolff transformations, we carried out a controlled derivation of quantum-mechanical low-energy theory for chiral fermions in Weyl semimetals with smooth TR and I breaking textures. The resulting effective Weyl Hamiltonian, describing carriers experiencing an effective curved spacetime, provides a unified approach also in the presence of strain. To illustrate the utility of the developed formalism, we proposed a concrete prescription to realize a spatial type I\textendash type II interface in magnetic topological insulators. This interface  is mathematically analogous to an event horizon of a black hole and may provide an experimental access to exotic high-energy phenomena. The developed low-energy theory is applicable to a wide variety of inhomogeneous Weyl semimetals and provides a powerful framework for analyzing and designing these systems.

An interesting avenue for future work is the generalization of our theory to time-dependent TR and I-breaking textures that give rise to non-stationary geometries and gauge fields. Another intriguing problem concerns the connections between different geometric responses in Weyl semimetals. As highlighted in the present work, low-energy carriers respond to magnetization through the change of effective geometry. This is analogous to elastic deformations in response to stress. Furthermore, thermal transport coefficients are also related to effective geometry  through gravitational response \cite{luttinger1964,2017Natur.547..324G}. These facts lead us to speculate on possible novel connections between seemingly distinct (magnetic, elastic, and thermal) response properties.

The authors acknowledge Aalto Center for Quantum Engineering for financial support.

\bibliography{Weyl.bib}

\widetext

\begin{center}
\Large SUPPLEMENTAL INFORMATION to ``Curved spacetime theory of inhomogeneous Weyl materials''
\end{center}

In this supplement we give detailed derivations of a number of results in the main text. To make the supplement self-contained we introduce the notation and background briefly. Our starting point is the parent Hamiltonian
\begin{eqnarray}
H_0=\kappa_i(\mathbf{k}) \gamma^i+m(\mathbf{k})\gamma_4.
\end{eqnarray}
 Throughout this paper we use the Minkowski metric with the signature $(-,+,+,+)$ and $\gamma_{i}=\gamma^i$ for $i=1,2,3$. Also note that the gamma matrices used in this paper are Hermitian, satisfying $\{\gamma_i,\gamma_j\}=2\delta_{ij}$ for $i,j=1,2,3,4$. The fifth gamma matrix is defined as $\gamma_5=\gamma_1\gamma_2\gamma_3\gamma_4$, which is also Hermitian and anticommutative with others. Using these gamma matrices we can construct ten additional matrices $\gamma_{ij}=-i[\gamma_i,\gamma_j]/2$, which can be separated into four groups as \cite{PhysRevB.84.235126,WeylMetamaterials_PRX2017} $\mathbf{b}=(\gamma_{23},\gamma_{31}, \gamma_{12})$, $\mathbf{b}'=(\gamma_{15},\gamma_{25},\gamma_{35})$, $\mathbf{p}=(\gamma_{14},\gamma_{24},\gamma_{34})$, and  $\varepsilon=\gamma_{45}$. It is straightforward to verify the commutation relations
 \begin{eqnarray}\label{Eq:su2}
  &&[b_i, b_j]=[b'_i, b'_j]=[p_i, p_j]=i2\varepsilon_{ijk}b_k,~[b_i,b'_i]=[b_i,p_i]=[b'_i,p_i]=0,
 \end{eqnarray}
 which will be useful later. 
  Using these matrices and the identity matrix, we can write the most general $4\times 4$ Hamiltonian.
 We require that $H_0$ is invariant under both time reversal and inversion, and the parameters $\kappa_i$ are odd function of $\mathbf{k}$ while $m$ is even. From these the transformation properties of the 15 matrices under time reversal and inversion can be determined: $\gamma_{1,2,3}$ are odd under both, $\gamma_4$ is even under both, $\mathbf{b}$ and $\mathbf{b}'$ are odd and time reversal but even under inversion, and $\mathbf{p}$ and $\varepsilon$ are even under time reversal but odd under inversion. 
 
 To obtain Weyl points, time-reversal or inversion symmetry must be broken. In the following we consider four types of spatially dependent symmetry breaking fields, $\mathbf{u}(\mathbf{r})\cdot\mathbf{b}$, $\mathbf{u}'(\mathbf{r})\cdot\mathbf{b}'$, $\mathbf{w}(\mathbf{r})\cdot\mathbf{p}$, and $f(\mathbf{r})\varepsilon$. It is clear that $\mathbf{u}(\mathbf{r})\cdot\mathbf{b}$ and $\mathbf{u}'(\mathbf{r})\cdot\mathbf{b}'$ always break the time reversal symmetry, and therefore we will call them the time reversal symmetry breaking terms. They can also break the inversion symmetry if the fields  $\mathbf{u}(\mathbf{r})$ and $\mathbf{u}'(\mathbf{r})$ are not even functions of $\mathbf{r}$. The other two terms, $\mathbf{w}(\mathbf{r})\cdot\mathbf{p}$ and $f(\mathbf{r})\varepsilon$, preserve the time reversal symmetry, and they  break the inversion symmetry if $\mathbf{w}(\mathbf{r})$ and $f(\mathbf{r})$ are not odd functions of $\mathbf{r}$. In the following we assume $\mathbf{w}(\mathbf{r})$ and $f(\mathbf{r})$ are not odd functions of $\mathbf{r}$ and call $\mathbf{w}(\mathbf{r})\cdot\mathbf{p}$ and $f(\mathbf{r})\varepsilon$ the inversion symmetry breaking terms.

\section{Effective Weyl Hamiltonians for TR-breaking cases}

In this section we present detailed derivations of the effective two-band Weyl Hamiltonians from the $4\times 4$ systems. 
\subsection{$\mathbf{u}(\mathbf{r})\cdot\mathbf{b}$ term} 
We first consider the Hamiltonian with the $\mathbf{u}\cdot \mathbf{b}$ term
\begin{eqnarray}
H=\kappa_i(\mathbf{k}) \gamma^i+m(\mathbf{k})\gamma_4 + \mathbf{u}(\mathbf{r})\cdot\mathbf{b}.
\end{eqnarray}
We parameterize the symmetry breaking fields as
\begin{eqnarray}
\mathbf{u}(\mathbf{r})=u(\mathbf{r})(\sin{\theta(\mathbf{r})}\cos{\phi(\mathbf{r})}, \sin{\theta(\mathbf{r})}\sin{\phi(\mathbf{r})}, \cos{\theta(\mathbf{r})}),
\end{eqnarray}
then $\mathbf{u}(\mathbf{r})\cdot\mathbf{b}$ can be rotated along the $z$ direction by a unitary transformation 
\begin{eqnarray}
W(\mathbf{r})=e^{-i\frac{\phi}{2}b_3}e^{-i\frac{\theta}{2}b_2},
\end{eqnarray}
i.e., 
 $W^\dag(\mathbf{r}) \mathbf{u}\cdot\mathbf{b} W(\mathbf{r})=u(\mathbf{r}) b_3$. The Hamiltonian after the transformation becomes
 \begin{eqnarray}
 H'&=&W^\dag H W=\frac{1}{2}W^\dag \kappa_i(\mathbf{k})W W^\dag \gamma^i W+ \frac{1}{2}W^\dag \gamma^i W W^\dag \kappa_i(\mathbf{k})W +W^\dag m(\mathbf{k}) W\gamma_4 + u(\mathbf{r}) b_3.
 \end{eqnarray}
In the above equation have used the facts that $[\gamma_4,W]=0$ and $[\kappa_i(\mathbf{k}), \gamma^i]=0$. Introducing the notations $W^\dag \gamma^i W=E^i_{~a}\gamma^a$ and $iW^\dag \partial_{\mathbf{r}} W=\bm{\omega}$, $H'$ can  be written as
\begin{eqnarray}\label{Eq:Dirac_TRB}
 H'&=&\frac{1}{2}\{\kappa_i(\mathbf{k} -\bm \omega), E^i_{~a}\gamma^a\}+ m(\mathbf{k}-\bm{\omega}) \gamma_4 + u(\mathbf{r}) b_3.
\end{eqnarray}
Explicitly,  we have%$E^{i}_{~a}$ and $\bm{\omega}$ are
\begin{eqnarray}
E^i_{~a}&=&\left[\begin{array}{ccc}
 \cos{\theta}\cos{\phi} & - \sin{\phi} & \cos{\phi}\sin{\theta} \\ 
\cos{\theta}\sin{\phi} & \cos{\phi}  & \sin{\phi}\sin{\theta}  \\ 
- \sin{\theta} & 0 & \cos{\theta}
\end{array} \right].\label{Eq:Eia}
\end{eqnarray}
Note that each column (row) of the matrix $E^i_{~a}$ is normalized and orthogonal to other columns (rows), i.e., $E^i_{~a}E^j_{~a}=\delta^{ij}$ and  $E^i_{~a}E^i_{~b}=\delta_{ab}$. 
Up to now what we have done is simply a local coordinate transformation: At every spatial point $\mathbf{r}_p$, we choose a local coordinate system in which the vector $\mathbf{u}(\mathbf{r}_p)$ is along the $z$ direction. 
It is easy to check that the new basis vector (frame) is $E_a=E^i_{~a}\partial_i$ (here $\partial_{i}$ is the basis vector in the tangent space, which in this case is simply the cartesian coordinate basis). Denoting the inverse matrix of $E^{i}_{~a}$ as $E_{i}^{~a}$, we have  $\partial_i = E_{i}^{~a}E_a$. Now it is clear that 
we can interpret $E^{i}_{~a}$ as the frame fields~\cite{Eguchi:1980jx}, and as we will see, $H'$ takes the same form as the Dirac Hamiltonian in curved space. 
 Using the frame fields we can define the metric $g^{ij}$ through the standard Euclidean metric $\delta^{ab}$ as $g^{ij}=E^{i}_{~a}E^{j}_{~b}\delta^{ab}=\delta^{ij}$. Note that $g^{ij}$ is still flat. With the metric $g_{ij}$ and its inverse $g^{ij}$, we can low or raise the spatial indices $i, j, k$, and the Euclidean metric $\delta_{ab}$ and $\delta^{ab}$ are used to low or raise the  indices $a, b, c$. For example, we can write $E_i^{~a}=g_{ij}\delta^{ab}E^j_{~b}$. 
We have $E_{i}^{~a}E^j_{~a}=\delta^i_j$ and $E_{i}^{~a}E^i_{~b}=\delta^a_b$, i.e., the inverse of the frame fields are obtained through lower or raise the index using the metric. 
 We may calculate $\bm{\omega}$ directly from its definition, but to understand its meaning we calculate it in the following way,
\begin{eqnarray}
i\partial_{r_j} E^{ia}\gamma_a&=&i\partial_{r_j} (W^\dag\gamma^i W)=i(\partial_{r_j} W^\dag)W W^\dag \gamma^i W+i W^\dag \gamma^i WW^\dag\partial_{r_j} W=-\omega_j E^{ia}\gamma_a+E^{ia}\gamma_a \omega_j, 
\end{eqnarray}
then we have 
\begin{eqnarray}
\gamma_a \omega_j-\omega_j \gamma_a=iE_{ia} \partial_{r_j} E^{ib}\gamma_b.\label{Eq:spinconnection} 
\end{eqnarray}
Since $W$ depends on $\gamma_{bc}$,  $\omega_j$ also depends on $\gamma_{bc}$ and can be written as $\omega_{j}=\omega_{j}^{~bc}\gamma_{bc}$. Because $[\gamma_a,\gamma_{bc}]=0$ if $a\ne b, c$ and $\{\gamma_a,\gamma_{bc}\}=0$ if $a=b$ or $a=c$, we have $\gamma^a\gamma_{bc} \gamma_a=-\gamma_{bc}$, and  therefore $\gamma^a\omega_j \gamma_a=-\omega_{j}$. Left multiplying both side of Eq.~\eqref{Eq:spinconnection} by $\gamma^a$ and summing over $a$, we find that $\bm{\omega}$ is the spin connection,
\begin{eqnarray}
\omega_j=\frac{i}{4}E_{i}^{~a} \partial_{r_j} E^{ib}\gamma_a\gamma_b=-\frac{1}{4}E_{i}^{~a} \partial_{r_j} E^{ib}\gamma_{ab}=-\frac{1}{4}E_{i}^{~a} \nabla_{r_j} E^{ib}\gamma_{ab}= \omega^{~ab}_{j}(\mathbf{r})\gamma_{ab}\equiv \omega^{~a}_{j}(\mathbf{r})b_a.
\end{eqnarray}
In the above derivations we have used that $\partial_{r_j}=\nabla_{r_j}$, which is a  consequence of $g_{ij}=\delta_{ij}$.
Taking $\kappa_i=k_i$, it is clear that $H'$ takes the form of the Dirac Hamiltonian in curved space~\cite{2011arXiv1106.2037Y}.  Explicitly, the spin connection reads
\begin{eqnarray}
\omega_i&=&\frac{\partial_{r_i} \phi}{2}\cos{\theta}\gamma_{12}
-\frac{\partial_{r_i} \phi}{2}\sin{\theta}\gamma_{23}+\frac{\partial_{r_i}\theta}{2}\gamma_{31}.
\end{eqnarray}

\subsection{Including strain effects}

It seems that we have just formulated the problem in a more complicated way; however, one advantage of Eq.~\eqref{Eq:Dirac_TRB} is that, it provides a unified description of inhomogeneous strain and spatially varying symmetry breaking fields. In the presence of small strain, the spatial metric is related to the strain tensor $u_{ij}$ as $g_{ij}=\delta_{ij}+2u_{ij}$, and the frame fields can be chosen as \cite{PhysRevLett.108.227205} $\bar{E}^{i}_{~a}=\delta^{i}_a-\delta_{aj}u^{ij}$, using the frame fields $\bar{E}^{i}_{~a}$ and the corresponding spin connection $\bar{\bm{\omega}}$, the Hamiltonian in the presence of strain and $\mathbf{u}\cdot\mathbf{b}$ fields is (we assume $\kappa_i=k_i$ and $m(\mathbf{k})=m$)
\begin{eqnarray}
H=\frac{1}{2}\left\{\bar{E}^i_{~a} \gamma^a, k_i -\bar{\omega}_i \right\}+m\gamma_4 + \mathbf{u}(\mathbf{r})\cdot\mathbf{b}.
\end{eqnarray}
After a unitary transformation we get
\begin{eqnarray}
H'=W^\dag H W =\frac{1}{2}\left\{\bar{E}^i_{~a} E^a_{~b}  \gamma^b, k_i -W^\dag \bar{\omega}_i W-\omega_i \right\}+m\gamma_4 + u(\mathbf{r})b_3,
\end{eqnarray}
as we have discussed above, physically, what we have done is  simply to rewrite $H$ by using the new frame fields $\tilde{E}^i_{~b}=\bar{E}^i_{~a} E^a_{~b}$.  
And it can be shown that $\tilde{\bm{\omega}}=W^\dag \bar{\bm{\omega}} W+\bm{\omega}$ is indeed the spin connection corresponding to the new frame fields,
\begin{eqnarray}
\tilde{\omega}_i&=&-\frac{\gamma_{ab}}{4}\tilde{E}_j^{~a}\nabla_i \tilde{E}^{jb}=-\frac{\gamma_{ab}}{4}\bar{E}_j^{~c}E_c^{~a}\nabla_i (\bar{E}^{je}E_{e}^{~b}),\\
&=&-\frac{\gamma_{ab}}{4}\left(\bar{E}_j^{~c}E_c^{~a}\partial_i( \bar{E}^{je}E_{e}^{~b})
+\tilde{E}_j^{~a}\Gamma^{j}_{im} \tilde{E}^{mb}\right),\\
&=&-\frac{\gamma_{ab}}{4}\left(E_c^{~a} E_{e}^{~b}\bar{E}_j^{~c}\partial_i  \bar{E}^{je}
+\bar{E}_j^{~c}E_c^{~a}\Gamma^{j}_{im} \bar{E}^{me}E_{e}^{~b} + \bar{E}_j^{~c} \bar{E}^{je} E_c^{~a}\partial_i E_{e}^{~b}\right),\\
&=&-\frac{\gamma_{ab}}{4}\left(E_c^{~a} E_{e}^{~b}\bar{\omega}_{i}^{~ce} +\omega_i^{~ab}\right)=W^\dag \bar{\omega}_i W+\omega_i.
\end{eqnarray}
In the following we shall understand Eq.~\eqref{Eq:Dirac_TRB} in the general sense that the frame fields in Eq.~\eqref{Eq:Eia} may also receive the above stated modification accounting for strain.

\subsection{Block diagonalization by the Schrieffer-Wolff transformation}

Another advantage is that $H'$ does not contain $b_1$ and $b_2$, making it more easier to be block-diagonalized. Assuming that $\mathbf{u}$ is independent of $\mathbf{r}$, then the frame fields are also independent of $\mathbf{r}$ and the spin connection vanishes. In this case $H'$ can be block-diagonalized by a momentum dependent unitary transformation $U(\mathbf{k})=e^{i\varphi \gamma_{34}/2}$ with $\tan{\varphi}=-m(\mathbf{k})/(\kappa_i(\mathbf{k}) E^i_{~3})$, 
\begin{eqnarray}\label{Eq:Weyl_uniform}
H'_{\mathrm{c}}=U^\dag(\mathbf{k}) H' U(\mathbf{k})=\kappa_i(\mathbf{k}) E^i_{~1}\gamma^1 +\kappa_i(\mathbf{k}) E^i_{~2}\gamma^2 +\sqrt{m^2(\mathbf{k})+(\kappa_i(\mathbf{k}) E^i_{~3})^2}\gamma^3+u\gamma_{12},
\end{eqnarray}
and the Weyl points are determined by $\kappa_i(\mathbf{k}) E^i_{~1}=\kappa_i(\mathbf{k}) E^i_{~2}=0$ and $\sqrt{m^2(\mathbf{k})+(\kappa_i(\mathbf{k}) E^i_{~3})^2}- u=0$. 
We can choose different unitary transformations and the resultant Hamiltonians take different forms, but conditions for the Weyl points are the same. 
In Ref.~\onlinecite{WeylMetamaterials_PRX2017}, the spatially varying $\mathbf{u}$ was investigated and the spin connection was ignored and the momentum and position operators are treated as $c$ numbers. Under these approximations, the effective  Weyl Hamiltonian takes the same form as Eq.~\eqref{Eq:Weyl_uniform}, with $E^i_{~a}$ being replaced by the position dependent counterpart. For this reason we call Eq.~\eqref{Eq:Weyl_uniform} the classical Hamiltonian, denoted by the subscript c. The purpose of this paper is to take into account the quantum nature of the momentum and position operators and include the effect of the spin connection. One may expect that there will be corrections to Eq.~\eqref{Eq:Weyl_uniform}, depending on the spin connection and derivatives of the frame fields. For slowly varying fields $\mathbf{u}(\mathbf{r})$, we expect the corrections are small and can be investigated in a controlled manner.
However,  Eq.~\eqref{Eq:Weyl_uniform} is not a good starting point to study the corrections because to obtain it, we have to use a complicated unitary transformation which is nonlinear in terms of $\bm{\kappa}$ and $E^{i}_{~a}$, making it difficult to include the corrections in a controlled way. A straightforward observation is that, for large $m$, we can expand $\sqrt{m^2+(\kappa_iE^i_{~3})^2}$ as
\begin{eqnarray}\label{Eq:Expansion}
\sqrt{m^2+(\kappa_iE^i_{~3})^2}=m+\frac{(\kappa_i E^i_{~3})^2}{2m}+\cdots,
\end{eqnarray}
and this result can also be obtained from $H'$ by a simple unitary transformation which allows us to have a better control of the corrections. Since we are mainly interested in the physics around the Weyl points, the large $m$ condition can be relaxed to $u(\mathbf{r})-|m(\mathbf{K}_W)|\ll u(\mathbf{r})$. Also note that if the opposite condition is satisfied, i.e., $|m(\mathbf{K}_W)|/u(\mathbf{r})\ll 1$, we can expand $\sqrt{m^2+(\kappa_iE^i_{~3})^2}$ as $\kappa_iE^i_{~3}+m^2/(2\kappa_iE^i_{~3})$, which can also be obtained from $H'$ through a simple unitary transformation.  
So our strategy is to  block-diagonalize $H'$ order by order by using the Schrieffer-Wolff transformation~\cite{Winkler}. This, of course, depends on whether there exits a control parameter like $m$ in Eq.~\eqref{Eq:Expansion}. In this sense our result is not as general as Eq.~\eqref{Eq:Weyl_uniform}; however, as we will show, the corrections from $\bm{\omega}$ can change the structure of the classical Hamiltonian and give rise to new phenomena that cannot be captured by Eq.~\eqref{Eq:Weyl_uniform}.

Before proceeding further, we briefly recall the Schrieffer-Wolff transformation~\cite{Winkler}.
Let $H$ be a Hamiltonian acting on the Hilbert space $\mathbb{H}$. Suppose $\mathbb{M}$ is a subspace of $\mathbb{H}$ we are interested in, then we can spit $H$ as $H=H_{\mathrm{diag}}+H_{\mathrm{off-diag}}$, where $H_{\mathrm{diag}}=P H P+(1-P)H(1-P)$ and  $H_{\mathrm{off-diag}}=P H (1-P)+(1-P)H P$ with $P$ being the projection operator that projects a state in $\mathbb{H}$ to $\mathbb{M}$. We use a unitary transformation $e^{-S}$ to eliminate the off-diagonal Hamiltonian,
\begin{eqnarray}
e^{-S} H e^{S}&=& H-[S,H]+\frac{1}{2}[S,[S,H]]+ \cdots\\
&=&H_{\mathrm{diag}}+H_{\mathrm{off- diag}}+[H_{\mathrm{diag}},S]+[H_{\mathrm{off- diag}},S]+\frac{1}{2}[S,[S, H_{\mathrm{diag}}] ]
+\frac{1}{2}[S,[S,H_{\mathrm{off- diag}} ]  ] + \cdots.
\end{eqnarray}
Let $S=S^{0}+S^{1}+\cdots$, and chose $S^0$ such that $H_{\mathrm{off- diag}}+[H_{\mathrm{diag}},S^0]=0$, then up to the lowest order, we get the effective Hamiltonian
\begin{eqnarray}
H_{\mathrm{eff}}=e^{-S^0} H e^{S^0}
&=&H_{\mathrm{diag}}+\frac{1}{2}[H_{\mathrm{off- diag}},S^0].
\end{eqnarray}
Note that $H_{\mathrm{eff}}$ is block-diagonalized because $S^0$ is off-diagonal. 

Now we apply this method to our problem. 
We are interested in the lowest order corrections, i.e., the corrections that contain only the first order derivatives with respect to $\mathbf{r}$ and (or) $\mathbf{k}$. To this end it is enough to expand $H'$ as
\begin{eqnarray}
H' 
&\approx&\frac{1}{2}\{\kappa_i(\mathbf{k})-\frac{\partial\kappa_i}{\partial k_j}\omega_j, E^i_{~a}\gamma^a\}+ m(\mathbf{k}) \gamma_4 +\frac{\partial m}{\partial k_j}\gamma_4\omega_j+ u(\mathbf{r}) b_3,\\
&=&\frac{1}{2}\{\kappa_i(\mathbf{k}), E^i_{~a}\gamma^a\}-\frac{1}{2}\frac{\partial\kappa_i}{\partial k_j}\omega_j^{~bc}E^i_{~a} \{\gamma_{bc}, \gamma^a\}+ m(\mathbf{k}) \gamma_4 +\frac{\partial m}{\partial k_j}\omega_j^{~ab}\gamma_4\gamma_{ab}+ u(\mathbf{r}) b_3,\\
&=&\frac{1}{2}\{\kappa_i(\mathbf{k}), E^i_{~a}\}\gamma^a+\frac{\partial\kappa_i}{\partial k_j}\omega_j^{~a}E^i_{~a} \gamma_{45}+ m(\mathbf{k}) \gamma_4 +\frac{\partial m}{\partial k_j}\omega_j^{~a}\gamma_{a5}+ u(\mathbf{r}) b_3,\\
&\equiv& \tilde{\kappa}_a(\mathbf{k},\mathbf{r})\gamma^a+m(\mathbf{k})\gamma_4+u(\mathbf{r}) b_3+ u'_i(\mathbf{k},\mathbf{r})b'_i+f(\mathbf{k},\mathbf{r})\varepsilon.
\end{eqnarray}
We thus obtain Eq.~(5) in the main paper. 
In the above expansion we have neglected the commutators between $\bm{\omega}$ and $\partial_{\mathbf{k}}\kappa$ and $\partial_{\mathbf{k}}m$, which depend on higher derivatives of the symmetry breaking field $\mathbf{u}$. Note that there are additional terms proportional to $\mathbf{b}'=(\gamma_{15},\gamma_{25},\gamma_{35})$ and $\varepsilon=\gamma_{45}$. These terms are different from what we may add to the parent Hamiltonian $H_0$ since in general they depend also on the momentum. More importantly, 
for slowly varying $u(\mathbf{r})$, these additional terms are small because they depend on the derivatives of $u(\mathbf{r})$, so we can treat them as  perturbations. In contrast, we do not assume the symmetry breaking fields directly added to $H_0$ are small. 
Since $\mathbf{u}'(\mathbf{k},\mathbf{r})$ contains the first order derivative of $m(\mathbf{k})$ with respect to $\mathbf{k}$, it is odd under time reversal (odd function of $\mathbf{k}$) and therefore  $u'_i(\mathbf{k},\mathbf{r}) b'_i$ is even under time reversal. Similarly, $f(\mathbf{k},\mathbf{r})\varepsilon$ is also even under time reversal. Also, if the time reversal symmetry breaking fields $\mathbf{u}(\mathbf{r})$ is even (odd) under inversion, $u'_i(\mathbf{k},\mathbf{r}) b'_i$ and $f(\mathbf{k},\mathbf{r})\varepsilon$ will also be even (odd) under inversion. In other words, the newly generated terms do not break the original symmetry of the Hamiltonian $H$.
 
 We assume that $m$ is large, then to find  the Schrieffer-Wolff transformation, 
it is convenient to use the following presentations of the $\gamma$ matrices, 
$\gamma_1=\tau_0\otimes\sigma_x$, $\gamma_2=\tau_0\otimes\sigma_y$, $\gamma_3=\tau_x\otimes \sigma_z$, and $\gamma_4=\tau_z\otimes\sigma_z$.
The diagonal part of $H'$ is  $H_{\mathrm{diag}}=\tilde{\kappa}_1\gamma_1+\tilde{\kappa}_2\gamma_2+m\gamma_4+u b_3 +u'_3 b'_3$, and $S^0$ can be chosen as
\begin{eqnarray}
S&^0=&-\frac{i}{4}\left\{\frac{1}{m}, \tilde{\kappa}_3\right\}\tau_y\otimes I
-\frac{i}{4}\left\{\frac{1}{m}, f\right\}\tau_y\otimes\sigma_z
%
%
%-\frac{i}{4m} \left[\tilde{\kappa}_3\tau_y\otimes I +f\tau_y\otimes\sigma_z\right]
%
%- \left[\tilde{\kappa}_3\tau_y\otimes I + f \tau_y\otimes\sigma_z\right]\frac{i}{4m}\\
-\frac{i }{4}\left\{\frac{1}{u}, u'_1  \right\}\tau_y\otimes\sigma_x - \frac{i }{4}\left\{\frac{1}{u}, u'_2 \right\}\tau_y\otimes\sigma_y,\\
&=&\frac{i}{4}\left\{\frac{1}{m}, \tilde{\kappa}_3\right\}\gamma_{34}
-\frac{i}{4}\left\{\frac{1}{m}, f\right\}\gamma_{5}-\frac{i }{4}\left\{\frac{1}{u}, u'_1  \right\}\gamma_{25}+ \frac{i }{4}\left\{\frac{1}{u}, u'_2 \right\}\gamma_{15}.
\end{eqnarray}
In the second line we have rewritten $S^0$ in a representation independent way. We also symmetrized the coefficients to make $e^{-S^0}$ Hermitian. It is easy to verify that
\begin{eqnarray}
\left[H_{\mathrm{diag}},S^0\right]=-H_{\mathrm{off- diag}}+H^{(2)}_{\mathrm{off- diag}},%\mathrm{high~ order~ derivatives}.
\end{eqnarray}
where $H^{(2)}_{\mathrm{off-diag}}$ depends on higher-order derivatives of $\mathbf{u}(\mathbf{r})$ as specified below.
%Note that $[\frac{v_F}{2} \sum_{i=1,2}(w_{ij} k_i + k_i w_{ij})I\otimes\sigma_j,v_F w_{ij}a^{~j}_i \tau_y\otimes\sigma_z]$ is off-diagonal, but of higher order, so it can be neglected.

After the Schrieffer-Wolff transformation, we obtain the effective Weyl Hamiltonian (we use the convention $\sigma_a=\sigma^a$),
\begin{eqnarray}
H_W&\equiv&d_a\sigma^a=  \sum_{i=1,2}\left(\tilde{\kappa}_i -\frac{f u'_i}{2u} -\frac{f u'_i}{2m}\right)\sigma_i +\left(-m+u-\frac{\tilde{\kappa}^2_3 +f^2 }{2m}+\frac{u^{'2}_1+u^{'2}_2}{2u}\right)\sigma_z-\left(\frac{f \tilde{\kappa}_3 }{m}+u'_3 \right) \sigma_0.\label{Eq:Weyl_TRB}
\end{eqnarray}
Note that the $k$ and $r$ dependent products should be understood as symmetrized, for example, $\frac{f u'_i}{u}=\frac{1}{4}(\frac{1}{u}u'_1+u'_i\frac{1}{u})f + \frac{1}{4}f (\frac{1}{u}u'_i+u'_i\frac{1}{u})$.
Since $f\sim \partial_{\mathbf{k}}\bm{\kappa} \partial_{\mathbf{r}}\mathbf{u}$ and $u'\sim
\partial_{\mathbf{k}}\bm{\kappa} \partial_{\mathbf{r}}\mathbf{u}$, 
the effective Hamiltonian is accurate up to the order of $(\partial_{\mathbf{k}} )^2 (\partial_{\mathbf{r}})^2$, and the next order correction is proportional to $\partial^2_{\mathbf{k}} (\partial_{\mathbf{r}})^2$, which comes from the second order expansions of $\bm{\kappa}(\mathbf{k}-\bm{\omega})$ and $m(\mathbf{k}-\bm{\omega})$. For the examples discussed in the main text we consider linear spectrum $\partial^2_{\mathbf{k}}\kappa=\partial^2_{\mathbf{k}}m=0$, so corrections actually vanish.  
If $f(\mathbf{k},\mathbf{r})$ and $\mathbf{u}'(\mathbf{k},\mathbf{r})$ are omitted, we get essentially the same result as the classical Hamiltonian Eq.~\eqref{Eq:Weyl_uniform}. The most important new ingredient in Eq.~\eqref{Eq:Weyl_TRB} is the $\sigma^0$ term, which is absent in  Eq.~\eqref{Eq:Weyl_uniform} and can lead to qualitatively different behaviour of the Weyl Hamiltonian. The Weyl points are determined by treating $\mathbf{k}$ and $\mathbf{r}$ as c numbers and solving $d_1=d_2=d_3=0$. We expand Eq.~\eqref{Eq:Weyl_TRB} around the Weyl point $\mathbf{K}_W$, 
\begin{eqnarray}\label{Eq:Weyl}
H_W\approx \left(d_0(\mathbf{K}_W)-\frac{ie^{~b}_0\partial_j e^j_{~b}}{2} \right)\sigma^0+ e^{i}_{~a}\left(k_i-K_{W,i}-\frac{i}{2}e^{~b}_{i}\partial_{j}e^{j}_{~b}\right)\sigma^a,
\end{eqnarray}
where 
\begin{eqnarray}
e^i_{~a}(\mathbf{r})=\frac{\partial d_a}{\partial k_i}\bigg|_{\mathbf{K}_W}
\end{eqnarray}
is the frame field. We introduce $e^{0}_{~0}=-1$ and $e^{0}_{~1,2,3}=0$ and then the frame fields can be written as a $4\times 4$ matrix. The Weyl points can be interpreted as the emergent $U(1)$ vector potential and $d_0(\mathbf{K}_W)$ is the scalar potential, both acquire corrections from the frame fields, ensuring that the Hamiltonian is Hermitian.
Using the frame fields and the Minkowski metric $\eta=\mathrm{diag}(-1,1,1,1)$, we obtain the emergent metric $g^{\mu\nu}=e^{\mu}_{~a}e^{\nu}_{b}\eta^{ab}$. Explicitly, $g^{00}=-1$, $g^{0i}=e^{i}_{~0}$, $g^{ij}=-e^i_{~0}e^j_{~0}+e^i_{~a}e^j_{~a}$. The component $g^{0i}$ is in general nonzero. Different from the frame fields in Eq.~\eqref{Eq:Dirac_TRB}, $e^\mu_{~a}(\mathbf{r})$ indeed leads to a curved spacetime. 
In general $e^{i}_{~0}$ is nozero, so the Weyl cone will be tilted.
 If $|e^{i}_{~0}e^{~l}_i|<1$, it is still a type I Weyl semimetal; if $|e^{i}_{~0}e^{~l}_i|>1$ the cone is over-tilted and we obtain a type II Weyl semimetal. 
 As pointed out by Volovik~\cite{2016JETPL.104..645V}, the interface between the type I and type II regions can be viewed as the event horizon of a black hole.
 As we demonstrate below, our theory suggest a concrete way to realize and control the event horizon.

\subsection{TR-breaking $\mathbf{u}'(\mathbf{r})\cdot\mathbf{b}'$ term} 
Now we discuss the time reversal symmetry breaking by $\mathbf{u}'(\mathbf{r})\cdot \mathbf{b}'$ term. As in the $\mathbf{u}\cdot\mathbf{b}$ case, we parameterize  $\mathbf{u}'(\mathbf{r})$ as
\begin{eqnarray}
\mathbf{u}'(\mathbf{r})=u'(\mathbf{r})(\sin{\theta(\mathbf{r})}\cos{\phi(\mathbf{r})}, \sin{\theta(\mathbf{r})}\sin{\phi(\mathbf{r})}, \cos{\theta(\mathbf{r})}),
\end{eqnarray}
then 
because of Eq.~\eqref{Eq:su2}, $\mathbf{u}'(\mathbf{r})\cdot \mathbf{b}'$  can be transformed to $u'(\mathbf{r})b'_3$
by applying the same unitary transformation $W(\mathbf{r})$.
 The Hamiltonian after the transformation becomes
\begin{eqnarray}
H'&=&W^\dag H W=\
\frac{1}{2}\{\kappa_i(\mathbf{k} -\bm \omega), E^i_{~a}\gamma^a\}+ m(\mathbf{k}-\bm{\omega}) \gamma_4 + u'(\mathbf{r}) b'_3,\\
&\approx & \tilde{\kappa}_a(\mathbf{k},\mathbf{r})\gamma^a+m(\mathbf{k})\gamma_4+u'(\mathbf{r}) b'_3+ u'_i(\mathbf{k},\mathbf{r})b'_i+f(\mathbf{k},\mathbf{r})\varepsilon.
\end{eqnarray}
Treating $\mathbf{k}$ and $\mathbf{r}$ as $c$ numbers and omitting $\bm{\omega}$, we can find a unitary transformation to block-diagonalize $H'$ and the resultant classical Hamiltonian can be written as,
\begin{eqnarray}
H'_{\mathrm{c}} = \tilde{\kappa}_3\gamma_3+ \sqrt{\tilde{\kappa}^2_1+\tilde{\kappa}^2_2+m^2}\gamma_4 + u'\gamma_{35}.
\end{eqnarray} 
Using the presentations of $\gamma$ matrices $\gamma_1=\tau_x\otimes \sigma_z$, $\gamma_2=\tau_y\otimes \sigma_z$, $\gamma_3=\tau_z\otimes \sigma_z$, and $\gamma_4=\tau_0\otimes \sigma_x$, we can write the Weyl Hamiltonian as
\begin{eqnarray}
H_W=-\tilde{\kappa}_3\sigma_z + \left(\sqrt{\tilde{\kappa}^2_1+\tilde{\kappa}^2_2+m^2}-u'\right)\sigma_x,
\end{eqnarray}
which actually describes a nodal line semimetal because the codimension of the degeneracy is 2. 
Taking into account $\bm{\omega}$, $H'$ can be block-diagonalized by 
\begin{eqnarray}
S^0&=&
 \frac{i}{4}\left\{ \frac{1}{m},\tilde{\kappa}_1 \right\} \tau_x\otimes\sigma_y
+\frac{i}{4}\left\{ \frac{1}{m},\tilde{\kappa}_2 \right\} \tau_y\otimes\sigma_y
-\frac{i}{4}\left\{ \frac{1}{u'+u'_3},u'_1 \right\} \tau_y\otimes\sigma_0
+\frac{i}{4}\left\{ \frac{1}{u'+u'_3},u'_2 \right\} \tau_x\otimes\sigma_0,\\
&=&
-\frac{i}{4}\left\{ \frac{1}{m},\tilde{\kappa}_1 \right\} \gamma_{14} 
-\frac{i}{4}\left\{ \frac{1}{m},\tilde{\kappa}_2 \right\} \gamma_{24}
-\frac{i}{4}\left\{ \frac{1}{u'+u'_3},u'_1 \right\} \gamma_{31}
+\frac{i}{4}\left\{ \frac{1}{u'+u'_3},u'_2 \right\} \gamma_{23}.
\end{eqnarray}
And the effective two band Hamiltonian is
\begin{eqnarray}
H_W=-\left(\tilde{\kappa}_3+f +\frac{\tilde{\kappa}_1u'_1+\tilde{\kappa}_2u'_2}{2(u'+u'_3)}
+\frac{\tilde{\kappa}_1u'_1+\tilde{\kappa}_2u'_2}{2m}
\right)\sigma_z + \left(m+\frac{\tilde{\kappa}^2_1+\tilde{\kappa}^2_2}{2m}-u'-u'_3
-\frac{u^{'2}_1+u^{'2}_2}{2(u'+u'_3)}
\right)\sigma_x,
\end{eqnarray}
with the codimension unchanged.

\section{Inversion symmetry breaking case} 
\subsection{$\mathbf{w}(\mathbf{r})\cdot\mathbf{p}$ term} 
In this section, we study the inversion symmetry breaking terms. We first consider the $\mathbf{w}\cdot\mathbf{p}$ term, and as before, the $\mathbf{w}$ field is parameterized as
\begin{eqnarray}
\mathbf{w}(\mathbf{r})=w(\mathbf{r})(\sin{\theta(\mathbf{r})}\cos{\phi(\mathbf{r})}, \sin{\theta(\mathbf{r})}\sin{\phi(\mathbf{r})}, \cos{\theta(\mathbf{r})}),
\end{eqnarray}
 and then because of Eq.~\eqref{Eq:su2}, we can use the same unitary transformation $W$ to change the Hamiltonian to
\begin{eqnarray}
H'&=&W^\dag H W=\
\frac{1}{2}\{\kappa_i(\mathbf{k} -\bm \omega), E^i_{~a}\gamma^a\}+ m(\mathbf{k}-\bm{\omega}) \gamma_4 + w(\mathbf{r}) \gamma_{34},\\
&\approx & \tilde{\kappa}_a(\mathbf{k},\mathbf{r})\gamma^a+m(\mathbf{k})\gamma_4+w(\mathbf{r}) \gamma_{34}+ u'_i(\mathbf{k},\mathbf{r})b'_i+f(\mathbf{k},\mathbf{r})\varepsilon.
\end{eqnarray} 
As we have discussed before, the additional terms $u'_i(\mathbf{k},\mathbf{r})b'_i$ and $f(\mathbf{k},\mathbf{r})\varepsilon$ preserve the time reversal symmetry. These terms also preserve the inversion symmetry if $\mathbf{w}(\mathbf{r})$ is even under inversion. If the $\mathbf{w}$ field contains a odd part, the additional terms also break the inversion symmetry.  (Recall that $\mathbf{w}$ must contain an even part for $\mathbf{w}\cdot\mathbf{p}$ to break the inversion symmetry.) So the newly generated terms do not break the original symmetry of $H$. 

The classical Hamiltonian is obtained by omitting $u'_i(\mathbf{k},\mathbf{r})b'_i$, $f(\mathbf{k},\mathbf{r})\varepsilon$ and the noncommutativity of $\mathbf{k}$ and $\mathbf{r}$, and can be written as
 \begin{eqnarray}
 H'_\mathrm{c}=\sqrt{\tilde{\kappa}^2_1 +\tilde{\kappa}^2_2}\gamma_1 +\tilde{\kappa}_3\gamma_3 +m\gamma_4 +w\gamma_{34}.
 \end{eqnarray}
 The Weyl points are determined by $\sqrt{\tilde{\kappa}^2_1 +\tilde{\kappa}^2_2}-w=0$, $\tilde{\kappa}_3=0$, and $m=0$.
 From these we see that, different from the time reversal symmetry breaking cases,  $m$ cannot be used as the controlling parameter in the inversion symmetry breaking case because it must vanish at the Weyl point (it may get small corrections as will be shown later, but this does change the conclusion). Instead, we may use $\tilde{\kappa}_1$ or $\tilde{\kappa}_2$ as the controlling parameter in the Schrieffer-Wolff transformation. Assuming $\tilde{\kappa}_1$ is large enough and using the representations of $\gamma$ matrices $\gamma_1=\tau_0\otimes\sigma_x$, $\gamma_2=\tau_x\otimes \sigma_y$, $\gamma_3=\tau_0\otimes\sigma_z$, and $\gamma_4=\tau_z\otimes\sigma_y$,  the diagonal part of $H'$ is $H_{\mathrm{diag}}=\tilde{\kappa}_1\gamma_1+\tilde{\kappa}_3\gamma_3+m\gamma_4 +w\gamma_{34}+u'_2\gamma_{25}$, and $S^0$ can be chosen as
 \begin{eqnarray}
 S^0&=&
 -\frac{i}{4}\left\{ \frac{1}{\tilde{\kappa}_1}, \tilde{\kappa}_2 \right\} \tau_x\otimes\sigma_z
 +\frac{i}{4}\left\{ \frac{1}{\tilde{\kappa}_1}, u'_1 \right\}
 \tau_y\otimes\sigma_y
 -\frac{i}{4}\left\{ \frac{1}{w}, f \right\}
 \tau_y\otimes\sigma_x
 +\frac{i}{4}\left\{ \frac{1}{w}, u'_3 \right\}
 \tau_x\otimes \sigma_0,\\
 &=&
 -\frac{i}{4}\left\{ \frac{1}{\tilde{\kappa}_1}, \tilde{\kappa}_2 \right\}\gamma_{12}
  +\frac{i}{4}\left\{ \frac{1}{\tilde{\kappa}_1}, u'_1 \right\}\gamma_5
  +\frac{i}{4}\left\{ \frac{1}{w}, f \right\}\gamma_{35}
  -\frac{i}{4}\left\{ \frac{1}{w}, u'_3 \right\}\gamma_{45}.
 \end{eqnarray}
 The effective Weyl Hamiltonian is
 \begin{eqnarray}
 H_W&=&\left(\tilde{\kappa}_1+\frac{\tilde{\kappa}^2_2+u^{'2}_1}{2\tilde{\kappa}_1} -\frac{f^2+u^{'2}_3}{2w}-w \right)\sigma_x 
 +\left( m(\mathbf{k}) +\frac{u'_1 f}{2w} +\frac{u'_1 f}{2\tilde{\kappa}_1}\right)\sigma_y
 +\left(\tilde{\kappa}_3 +\frac{u'_3u'_1}{2\tilde{\kappa}_1}  -\frac{u'_3u'_1}{2w} \right)\sigma_z \\
 &&+ \left(u'_2-\frac{\tilde{\kappa}_2 u'_1}{\tilde{\kappa}_1} \right)\sigma_0,
 \end{eqnarray} 
 which takes the same form as Eq.~\eqref{Eq:Weyl_TRB}.
 
 \subsection{$f(\mathbf{r})\varepsilon $ term} 
Now let us consider the inversion symmetry breaking by $f(\mathbf{r})\varepsilon$. The Hamiltonian is
 \begin{eqnarray}
    H=\kappa_1(\mathbf{k})\gamma_1+\kappa_2(\mathbf{k})\gamma_2+\kappa_3(\mathbf{k})\gamma_3+m(\mathbf{k})\gamma_4+f(\mathbf{r})\gamma_{45}.
 \end{eqnarray}
 In this case we  parameterize $\bm{\kappa}$
 as
 \begin{eqnarray}
 \bm\kappa(\mathbf{k})=\kappa(\mathbf{k}) (\sin{\theta(\mathbf{k})}\cos{\phi(\mathbf{k})}, \sin{\theta(\mathbf{k})}\sin{\phi(\mathbf{k})}, \cos{\theta(\mathbf{k})} ).
 \end{eqnarray}
 Using a momentum dependent unitary transformation $U(\mathbf{k})=e^{-i\frac{\phi(\mathbf{k})}{2}\gamma_{12}}e^{-i\frac{\theta (\mathbf{k}) }{2}\gamma_{31}}$, $\bm{\kappa}\cdot\bm{\gamma}$ can be rotated to $\kappa \gamma_3$.
 The Hamiltonian after the transformation becomes
 \begin{eqnarray}
 H'&=&U^\dag H U=\kappa(\mathbf{k})\gamma_3+m(\mathbf{k})\gamma_4+f(\mathbf{r}+iU^\dag \partial_\mathbf{k} U)\gamma_{45},\\
 &\approx& \kappa(\mathbf{k})\gamma_3+m(\mathbf{k})\gamma_4+f(\mathbf{r})\gamma_{45}+i\partial_{r_i}f(\mathbf{r})U^\dag \partial_\mathbf{k} U \gamma_{45},\\
 &=&\kappa(\mathbf{k})\gamma_3+m(\mathbf{k})\gamma_4+f(\mathbf{r})\gamma_{45}-\partial_{r_i}f(\mathbf{r})a_i^{~j} \gamma_{j}.
 \end{eqnarray}
In the last line we have introduced the notations
 \begin{eqnarray}
 a_i(\mathbf{r})&=&i U^\dag(\mathbf{r}) \partial_{k_i} U(\mathbf{r})=\frac{\partial_{k_i} \phi}{2}\cos{\theta}\gamma_{12}
 -\frac{\partial_{k_i} \phi}{2}\sin{\theta}\gamma_{23}+\frac{\partial_{k_i}\theta}{2}\gamma_{31}=a^{~j}_{i}(\mathbf{k})b_j.
 \end{eqnarray}
 Note that $a_i^{~3}$ is the Berry connection corresponding to the upper band of the Hamiltonian $\kappa_i(\mathbf{k})\gamma_i$, and the corresponding quantum metric can be written as $g_{ij}=2\Re (a_i^{~1}-i a_i^{~2})(a_j^{~1}+i a_j^{~2})$. To block-diagonalize $H'$, it is convenient to use the representation of the $\gamma$ matrices $\gamma_1=\tau_x\otimes \sigma_z$, $\gamma_2=\tau_y\otimes \sigma_z$, $\gamma_3=\tau_z\otimes \sigma_z$, and $\gamma_4=\tau_0\otimes \sigma_x$, and then after a Schrieffer-Wolff transformation, we obtain the effective Hamiltonian as
\begin{eqnarray}\label{Eq:ISB_e}
H_W=\left(f(\mathbf{r})-\kappa(\mathbf{k})+\partial_{r_i}f(\mathbf{r})a_i^{~3} +\frac{\partial_{r_i}f(\mathbf{r}) \partial_{r_j}f(\mathbf{r}) g_{ij} }{4 \kappa}\right)\sigma_z +m(\mathbf{k})\sigma_x,
\end{eqnarray}
which in general describes a nodal line  semimetal instead of a Weyl semimetal. 

 \section{Realization of spatial interface between type I and type II Weyl fermions}
 In this section we demonstrate that the type I and type II Weyl fermion can be realized and the interface between them can be tuned.
 We consider a simple Hamiltonian
 \begin{eqnarray}
 H=k_i\gamma_i + m\gamma_4+\mathbf{u}(\mathbf{r})\cdot \mathbf{b},
 \end{eqnarray}
 with TR-breaking triplet
 \begin{eqnarray}
\mathbf{u}(\mathbf{r})=u(\mathbf{r})(\sin{\theta(\mathbf{r})}\cos{\phi(\mathbf{r})}, \sin{\theta(\mathbf{r})}\sin{\phi(\mathbf{r})}, \cos{\theta(\mathbf{r})}),
\end{eqnarray}
 which describes real materials such as 3d topological insulators with magnetic texture  or topological insulator-magnet heterostructures (as discussed in Ref.~\onlinecite{WeylMetamaterials_PRX2017}). After the unitary transformation which brings $\mathbf{u}$ parallel to $z$ axis, the Hamiltonian becomes
 \begin{eqnarray}
 H'&=&W^\dag H W=
 \tilde{\kappa}_a(\mathbf{k},\mathbf{r})\gamma^a+m\gamma_4+u(\mathbf{r}) b_3+f(\mathbf{k},\mathbf{r})\varepsilon,
 \end{eqnarray}
 where 
 \begin{eqnarray}
 \tilde{\kappa}_1&=&\cos{\theta}\cos{\phi} k_x+\cos{\theta}\sin{\phi} k_y-\sin{\theta}k_z,\\
 \tilde{\kappa}_2&=&-\sin{\phi}k_x+\cos{\phi}k_y,\\
 \tilde{\kappa}_3&=&\cos{\phi}\sin{\theta}k_x+\sin{\phi}\sin{\theta}k_y+\cos{\theta}k_z,\\
f&=&\frac{\partial_z\phi + \cos{\phi}\partial_y\theta-\sin{\phi}\partial_x\theta}{2}
 \end{eqnarray}
 (the products in $\bm{\tilde{\kappa}}$ should be understood as symmetrized).
 The effective Weyl Hamiltonian reads 
 \begin{eqnarray}
 H_W&=&  \sum_{i=1,2} \tilde{\kappa}_i \sigma_i +\left(-m+u-\frac{\tilde{\kappa}^2_3 +f^2 }{2m}\right)\sigma_z-\left(\frac{f \tilde{\kappa}_3 }{m} \right) \sigma^0,
% \approx e^i_{~a}(k_i-K_{W,i})\sigma^a + V,
 \end{eqnarray}
 and the Weyl points are $\pm \mathbf{K}_W=\pm K_W(\sin{\theta}\cos{\phi}, \sin{\theta}\sin{\phi},\cos{\theta})$ with $K_W=\sqrt{2m(u-m)-f^2}$. We assume that $2m(u-m)>f^2$ such that there are always two separated Weyl points. As we have mentioned before, for the first order Schrieffer-Wolff transformation to be a good approximation, the condition $u-m\ll m$ should be satisfied.
 Expanding around $\mathbf{K}_W$, we find (the corrections to the effective gauge fields have been omitted)
 \begin{eqnarray}
H_W \approx e^i_{~a}(k_i-K_{W,i})\sigma^a -\frac{fK_W}{m},
 \end{eqnarray}
 with the frame and coframe fields being
 \begin{eqnarray}
 e^\mu_{~a}=\left[\begin{array}{cccc}
 -1 & 0 & 0 & 0 \\ 
 -\frac{f}{m}\cos{\phi}\sin{\theta} & \cos{\theta}\cos{\phi} & -\sin{\phi} & -\frac{K_W}{m}\cos{\phi}\sin{\theta}\\ 
 -\frac{f}{m}\sin{\phi}\sin{\theta} & \cos{\theta}\sin{\phi} & \cos{\phi}  & -\frac{K_W}{m}\sin{\phi}\sin{\theta}\\
 -\frac{f}{m}\cos{\theta} & -\sin{\theta} & 0 & -\frac{K_W}{m}\cos{\theta}
 \end{array} \right],
 e_\mu^{~a}=\left[\begin{array}{cccc}
 -1 & 0 & 0 & \frac{f}{K_W} \\ 
 0 & \cos{\theta}\cos{\phi} & -\sin{\phi} & -\frac{m}{K_W}\cos{\phi}\sin{\theta}\\ 
 0 & \cos{\theta}\sin{\phi} & \cos{\phi}  & -\frac{m}{K_W}\sin{\phi}\sin{\theta}\\
 0 & -\sin{\theta} & 0 & -\frac{m}{K_W}\cos{\theta}
 \end{array} \right],~~~~
 \end{eqnarray}
 Then it is straightforward to get $e^1_{~0}e_i^{~1}=e^1_{~0}e_i^{~2}=0$ and $e^1_{~0}e_i^{~3}=f/K_W$, so $|f/K_W|<1$ describes a type I Weyl point and $|f/K_W|>1$ describes a type II Weyl point. The interface between the type I and type II regions, or the event horizon, is  determined by $|f/K_W|=1$.
 
 From the frame fields we can calculate the emergent metric. 
 Note that $e^i_{~1}=E^i_{~1}$, $e^i_{~2}=E^i_{~2}$, and $e^i_{~3}=-K_W E^i_{~3}/m $, so in terms of the  local spatial basis $E_a$  where $\mathbf{u}$ is always along the $z$ direction, the metric takes a particularly simple form,
 \begin{eqnarray} 
 g^{ab}=\left[\begin{array}{cccc}
 -1 & 0 & 0 & -\frac{f}{m}\\ 
 0 & 1 & 0 & 0\\  
 0 & 0 & 1 & 0\\ 
 -\frac{f}{m} & 0 & 0 & \frac{K^2_W-f^2}{m^2}\\ 
 \end{array} \right],~~ g_{ab}=\left[\begin{array}{cccc}
 -1+\frac{f^2}{K^2_W} & 0 & 0 & -\frac{f m}{K^2_W}\\ 
 0 & 1 & 0 & 0\\  
 0 & 0 & 1 & 0\\ 
 -\frac{f m}{K^2_W} & 0 & 0 & \frac{m^2}{K^2_W}\\ 
 \end{array} \right].
 \end{eqnarray}
 The metric is analogous to the Schwarzschild metric in the Gullstrand-Painlev\'{e} coordinates~\cite{2016JETPL.104..645V,PhysRevLett.46.1351,PhysRevD.51.2827} when replacing $f/K_W$ by the velocity of a freely falling observable $v(r)$. Thus, it is clear that the even horizon is determined by  $|f/K_W|=1$, consistent with the result obtained from the dispersion of the Weyl fermions.
 
 %\begin{figure}
 %    \includegraphics[width=0.9\textwidth]{supplement.png}
% 	\caption{The interfaces between type I and type II Weyl fermions corresponding to different magnetic textures. (a), $\mathbf{u}=u (\sin{r/\xi},0,\cos{r/\xi})$ with  $u/m=11/10$ and $m\xi=\sqrt{2}$. The interface is a conical surface defined by $x^2+z^2=y^2/4$. 
% 	(b), $\mathbf{u}=(-u_{||}y/\sqrt{x^2+y^2},u_{||}x/\sqrt{x^2+y^2} ,u_3)$, where $u_{||}=u_0 \exp{(-\sqrt{x^2+y^2}/\xi)}$, $u_3/m=10/9$, $u_0/u_3=1/9$, and $m\xi=9/70$. The interface is a cylinder surface whose radius is about $0.232\xi$.
% 	The values of the coordinates in both plots are given in the units of $\xi$.}\label{Fig:interface}
% \end{figure}
 
 Now we study several magnetization configurations in detail.  Ref.~\onlinecite{WeylMetamaterials_PRX2017} studied a Skyrmion like magnetic texture with $u=\mathrm{constant}$, $\cos{\phi}=x/\sqrt{x^2+y^2}$,  $\sin{\phi}=y/\sqrt{x^2+y^2}$, and $\theta=\sqrt{x^2+y^2}/\xi$ . In this case $f= 0$, and therefore the Weyl point is untilted. To realize a type II Weyl point we can simply replace the Skyrmion like magnetic texture by a chiral one with $\cos{\phi}=-y/\sqrt{x^2+y^2}$ and  $\sin{\phi}=x/\sqrt{x^2+y^2}$. Then $f=-1/(2\xi)$ is a constant and $4\xi^2m(u-m)>1$ leads to a type II Weyl point.
  
 To realize the interface between type I and type II Weyl fermions, we consider another  magnetic texture with $u=\mathrm{constant}$, $\cos{\phi}=1$,  $\sin{\phi}=0$, and $\theta=\sqrt{x^2+y^2+z^2}/\xi$.  
 The symmetry breaking field is slowly varying if the length scale  we are interested in is smaller than $\xi$. 
 In this case we find $f=y/(2r\xi)$ and then $K_W=\sqrt{8m(u-m)r^2\xi^2-y^2}/(2r\xi)$. Since the minimal value of $|f|$ is zero, there always exists a type I region, and the interface between type I and type II regions is determined by  $y^2=4\xi^2m(u-m)r^2$, which defines a conical surface. Clearly, the position of the interface can be tuned by changing parameters such as $\xi$, which means that the event horizon may be tuned experimentally, providing a way to simulate the Hawking radiation~\cite{2016JETPL.104..645V}.

 As the last example, we consider the magnetic configuration~\cite{PhysRevB.87.235306} characterized by $u(r)=\sqrt{u^2_{||}(r)+u^2_3}$, $\cos{\phi}=-y/r$,  $\sin{\phi}=x/r$, and $u_3=u(r)\cos{\theta}$ is independent on $r=\sqrt{x^2+y^2}$. Correspondingly, $f= -u_3 \partial_{r} u_{||}/(2u^2)$. Physically, $u_3$ describes a uniform magnetization in the $z$ direction and $u_1$ and $u_2$ describes a chiral magnetization with a position dependent amplitude $u_{||}(\mathbf{r})$. It is reasonable to assume $u_{||}(\mathbf{r})\to 0$ when $r\to \infty$, and for simplicity  we  take $u_{||}(\mathbf{r})=u_{0}e^{-r/\xi}$. Then the horizon is determined by $4m(u-m)u^4\xi^2=u^2_3 u^2_{||}$, which defines a cylinder.

\end{document}